\documentclass[prd,twocolumn,nofootinbib,superscriptaddress]{revtex4-1}
\input epsf
\usepackage{graphics}
\usepackage{amsmath,amssymb}
\usepackage{color}
\usepackage{dcolumn}
\usepackage{hyphenat}
\usepackage{multirow}

\usepackage{graphicx}

\def\be{\begin{equation}}
\def\ee{\end{equation}}
\def\ba{\begin{eqnarray}}
\def\ea{\end{eqnarray}}

\newcommand{\beqa}{\begin{eqnarray}}
\newcommand{\eeqa}{\end{eqnarray}}

\newcommand{\beq}{\begin{equation}}
\newcommand{\eeq}{\end{equation}}

\newlength{\tskip}
\newlength{\colwidth}

\begin{document}

\title{Relieving the Hubble tension with primordial magnetic fields}

\author{Karsten Jedamzik} 
\affiliation{Laboratoire de Univers et Particules de Montpellier, UMR5299-CNRS, Universite de Montpellier, 34095 Montpellier, France}
\email[]{karsten.jedamzik@umontpellier.fr}

\author{Levon Pogosian} 
\affiliation{Department of Physics, Simon Fraser University, Burnaby, BC, V5A 1S6, Canada}
\affiliation{Institute of Cosmology and Gravitation, University of Portsmouth, Portsmouth, PO1 3FX, UK}
\email[]{levon@sfu.ca}

\begin{abstract}
The standard cosmological model determined from the accurate cosmic microwave background measurements made by the Planck satellite implies a value of the Hubble constant $H_0$ that is $4.2$ standard deviations lower than the one determined from Type Ia supernovae. The Planck best fit model also predicts higher values of the matter density fraction $\Omega_m$ and clustering amplitude $S_8$ compared to those obtained from the Dark Energy Survey Year 1 data. Here we show that accounting for the enhanced recombination rate due to additional small-scale inhomogeneities in the baryon density may solve both the $H_0$ and the $S_8-\Omega_m$ tensions. The additional baryon inhomogeneities can be induced by primordial magnetic fields present in the plasma prior to recombination. The required field strength to solve the Hubble tension is just what is needed to explain the existence of galactic, cluster, and extragalactic magnetic fields without relying on dynamo amplification. Our results show clear evidence for this effect and motivate further detailed studies of primordial magnetic fields, setting several well-defined targets for future observations.
\end{abstract}

\maketitle

The standard $\Lambda$ Cold Dark Matter ($\Lambda$CDM) model of cosmology has withstood two decades of testing against the ever improving observational data. However, with multiple independent types of measurements producing very accurate results over the past few years, some tensions between the $\Lambda$CDM parameters obtained from different datasets have emerged. Most notable of them is the discrepancy between the value of the Hubble constant $H_0$ inferred from the cosmic microwave background (CMB) measurements by Planck \cite{Aghanim:2018eyx} and the one obtained from type Ia supernovae (SNIa) and certain other types of measurements in the $z \sim 0.01-1$ redshift range. In particular, the Planck best fit value of $H_0=67.36 \pm 0.54$ km s$^{-1}$ Mpc$^{-1}$ \cite{Aghanim:2018eyx} agrees very well with $H_0=67.4^{+1.1}_{-1.2}$ km s$^{-1}$ Mpc$^{-1}$ obtained from the Dark Energy Survey Year 1 (DES-Y1) clustering and weak lensing data combined with Baryon Acoustic Oscillations (BAO) measurements from a variety of spectroscopic surveys \cite{Abbott:2017smn}. But it is significantly ($4.2\sigma$) lower than $H_0=73.5 \pm 1.4$ km s$^{-1}$ Mpc$^{-1}$ measured by the Supernovae, H0, for the Equation of State of Dark Energy (SH0ES) collaboration \cite{Reid:2019tiq} using SNIa luminosities calibrated on Cepheid variable stars. SNIa studies using alternative calibration methods also find higher values of $H_0$ \cite{2018AAS...23231902P,Freedman:2019jwv,Yuan:2019npk,Huang:2019yhh} (see \cite{Verde:2019ivm} for a discussion). Determinations of $H_0$ that do not rely on SNIa include the Megamaser Cosmology Project (MCP) \cite{Reid:2008nm} that obtained $H_0=73.9 \pm 3.0$ km s$^{-1}$ Mpc$^{-1}$ \cite{Pesce:2020xfe} from very-long-baseline interferometry observations of water masers orbiting supermassive black holes, and the H0 Lenses in COSMOGRAILÕs Wellspring (H0LiCOW) value of $H_0=73.3^{+1.7}_{-1.8}$ km s$^{-1}$ Mpc$^{-1}$ \cite{Wong:2019kwg} inferred from a joint analysis of six gravitationally lensed quasars with measured time delays.

Another, albeit somewhat weaker, tension exists between the values of the amplitude of galaxy clustering $S_8$ and the matter fraction $\Omega_m$ in the Planck best-fit model and those inferred from the latest surveys of large scale structure. Specifically, the Planck values are $S_8=0.832 \pm 0.013$ and $\Omega_m=0.3153 \pm 0.0073$ \cite{Aghanim:2018eyx}, while the DES-Y1 weak lensing and galaxy clustering data yields $S_8=0.783^{+0.021}_{-0.025}$ and $\Omega_m=0.264^{+0.032}_{-0.019}$ \cite{Abbott:2017wau}. 
   
Many extensions of the $\Lambda$CDM model have been proposed with the aim of resolving the $H_0$ problem (see \cite{Knox:2019rjx} for a review). Late-time dynamical dark energy or modifications of gravity can reduce the tension \cite{Zhao:2017cud,Wang:2018fng,Raveri:2019mxg,DiValentino:2019jae} but there is no evidence for them otherwise and, aside from the higher value of $H_0$, the dynamics of the universe in the $z \sim 0.1-1$ redshift range is in good agreement with $\Lambda$CDM. 
Importantly, a higher $H_0$ is preferred by all measurements that do not rely on our understanding of the recombination history and the determination of the sound horizon at the photon and the baryon decoupling epochs. If the sound horizon at recombination happened to be smaller due to a yet unaccounted effect, the observed angular acoustic scales in the CMB anisotropies and galaxy density fluctuations would imply a larger value of $H_0$. This would happen, for example, if the dark energy density became dynamically important for a period of time before recombination \cite{Poulin:2018cxd,Agrawal:2019lmo,Lin:2019qug,Sakstein:2019fmf}. However, as recently pointed out in \cite{Hill:2020osr}, such early dark energy (EDE) would delay the development of gravitational potentials, requiring a larger matter density to compensate and worsening the $S_8$-$\Omega_m$ tension. The situation may be better when considering modifications in the neutrino sector of the standard model \cite{Kreisch:2019yzn}.

\begin{figure}[htbp]
\centering
\includegraphics[width=0.48\textwidth]{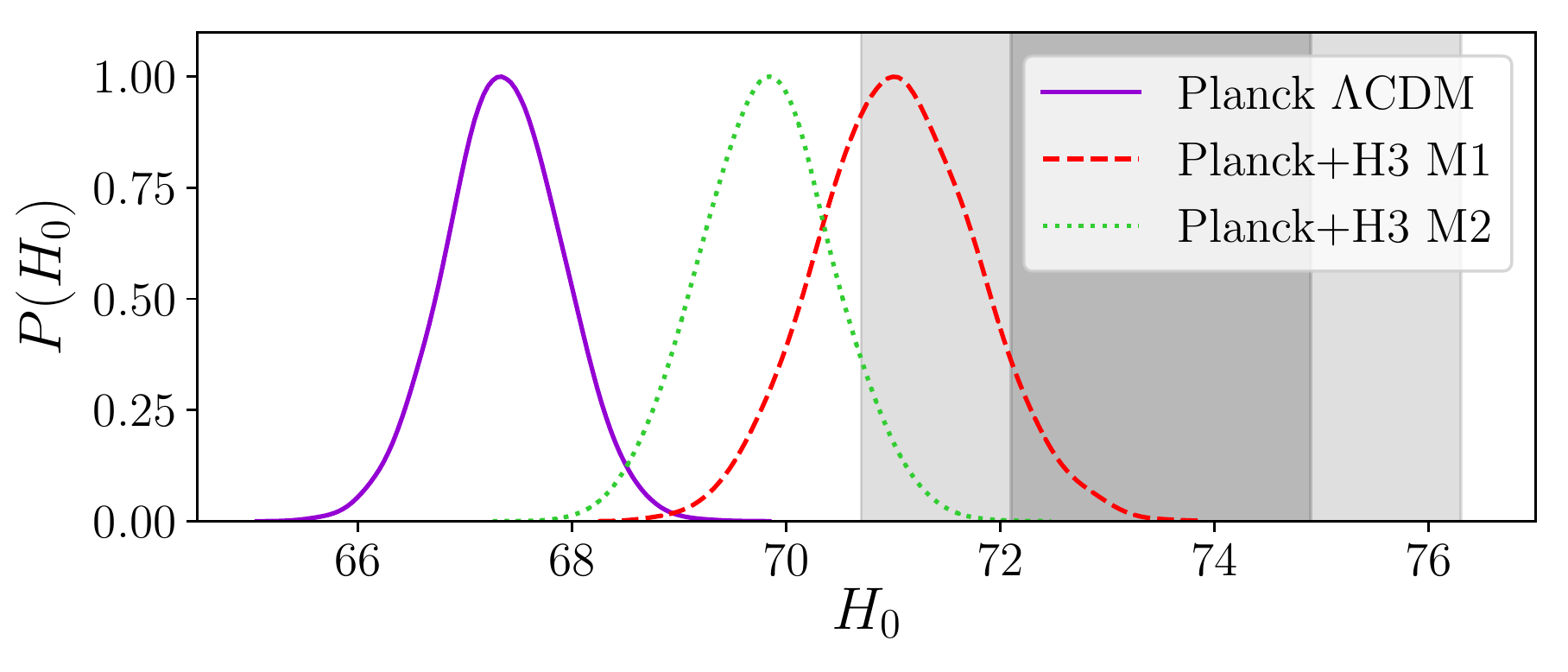}
\includegraphics[width=0.48\textwidth]{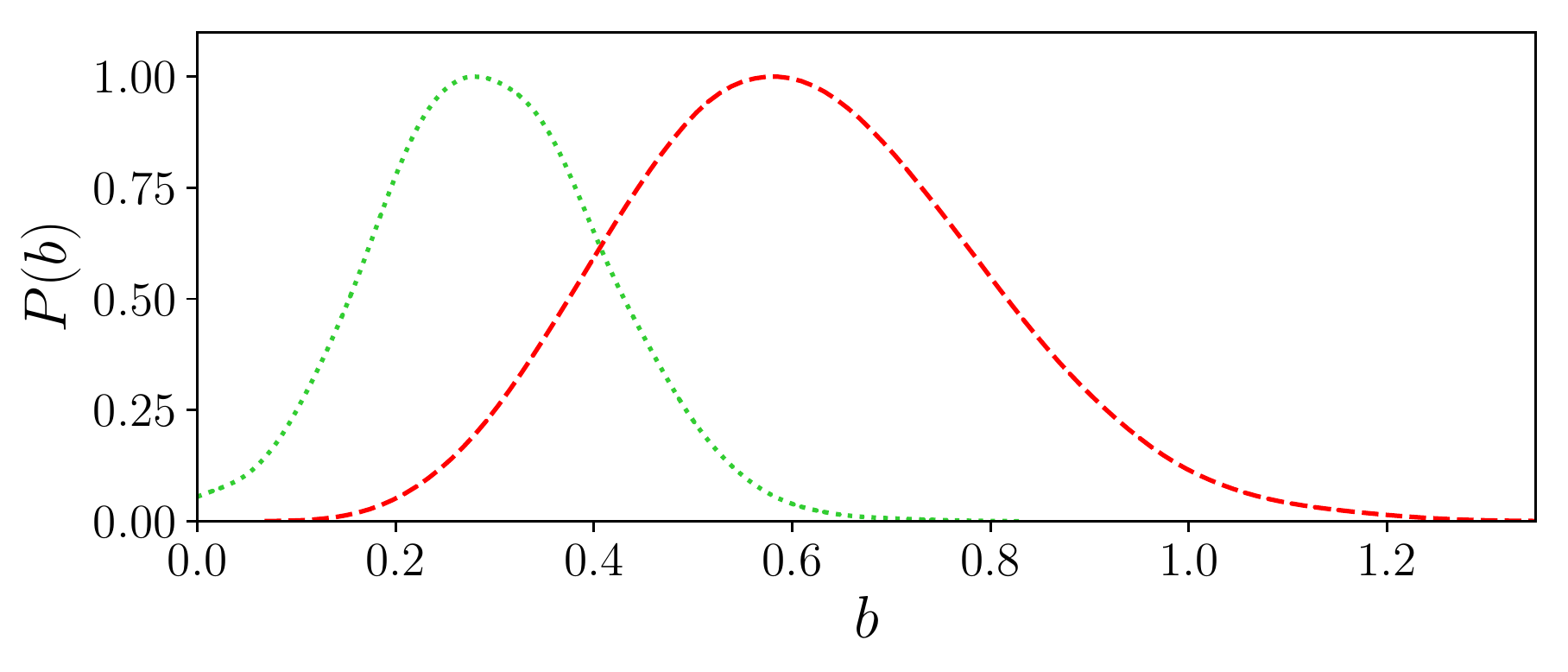}
\caption{\label{fig:clumping} The marginalized $H_0$ PDF for the Planck best fit $\Lambda$CDM model and the two baryon clumping models, M1 and M2, fit to Planck combined with H3. The bottom panel shows the PDF of the clumping parameter $b$. The shaded regions show the $68$\% and $95$\% CL of $H_0$ from SH0ES \cite{Reid:2019tiq}.}
\end{figure}

In this {\it Letter} we show that both the $H_0$ and the $S_8-\Omega_m$ tensions may be greatly alleviated when allowing for small-scale, mildly non-linear inhomogeneities in the baryon density shortly before recombination. Such baryon inhomogeneities are motivated by detailed studies of the evolution of primordial magnetic fields (PMFs) before recombination~\cite{Jedamzik:2013gua,Jedamzik:2018itu}. In a nutshell, on scales well below the photon mean free path (and the Silk damping scale) the effective speed of sound is far lower than that of a relativistic plasma such that PMFs of relatively moderate $\sim 0.1$ nano-Gauss (nG) strength\footnote{In this paper, the field strengths are quoted at their comoving redshifted present day values.} can generate baryon inhomogeneities on $\sim\,$kpc scales. Based on a comprehensive understanding of the PMF evolution in the early universe~\cite{Banerjee:2004df}, the effect has been derived analytically~\cite{Jedamzik:2013gua} and confirmed numerically~\cite{Jedamzik:2018itu}. In fact, the latter study used the Planck 2013 CMB data with a variety of other datasets to stringently constrain PMFs. 

Even though small-scale baryon inhomogeneities on $\sim\,$kpc scales do not directly source CMB temperature and polarization anisotropies, their existence impacts the observed CMB anisotropies by profoundly changing the ionization history, and therefore the epoch of recombination. The ionization fraction $\chi_e$ is determined by a balance between the recombination rate, proportional to the electron density square $n_e^2$, and the ionization rate proportional to the neutral hydrogen density $n_H$. As any inhomogeneous universe has $\langle n_e^2\rangle > \langle n_e\rangle^2 = \langle n_{e}^2\rangle|_{\rm homo}$, the average recombination rate is increased and the recombination occurs earlier in an inhomogeneous universe. An earlier recombination, in turn, reduces the sound horizon $r_*$ at recombination. The corresponding impact on the CMB anisotropy spectra would be a shift of the locations of all the acoustic peaks to smaller scales. Since the positions of the peaks, $\ell_p \propto r_{\rm ls}/r_*$, are measured with great accuracy by Planck, one would need a smaller conformal distance to last scattering  $r_{\rm ls}$ to compensate for the shift, which requires a larger $H_0$.

In Refs.~\cite{Jedamzik:2013gua,Jedamzik:2018itu} a baryon clumping factor $b = (\langle n_b^2\rangle - \langle n_b \rangle^2)/\langle n_b \rangle^2$ was introduced to gauge the amplitude of the PMF generated inhomogeneities. Since the recombination term is quadratic, one could naively expect that the reduction of the ionization fraction in inhomogeneous universes is fully determined by the average density and the clumping factor alone, {\it i.e.} the first two moments of the baryon density probability distribution function (PDF). However, a more careful analysis shows that the reduction of $\chi_e$ depends on all moments of the baryon density PDF. In the absence of detailed knowledge of the PDF, we employed two different PDFs implemented via a three-zone model described below.

We have modified the publicly available Code for Anisotropies in the Microwave Background (CAMB) \cite{Lewis:1999bs} to include the effects of small-scale baryon inhomogeneities on the recombination history. In particular, we make the code RECFAST \cite{Seager:1999bc,Seager:1999km,Wong:2007ym} calculate the evolution of $\chi_e$ independently in three different zones, with the electron density in each zone drawn from a PDF normalized to set values of $\langle n_b\rangle$ and $b$, and take the appropriate average. The independent-zone approximation is well-justified, since the $\sim 1\,$kpc length scale corresponding to the clumping effect is much bigger than the mixing scale of $\sim 1\,$pc (comoving) set by the diffusion length of baryons at recombination. Turbulent mixing due to MHD turbulence, on the other hand, is absent, since the plasma is in a low Reynolds number viscous state due to the strong residual photon-electron interactions.

Having three zones keeps the computational costs down, while still demonstrating the importance of accounting for the shape of the baryon density PDF. We have chosen two distinct distributions, hereafter referred to as Model 1 (M1) and Model 2 (M2), detailed in the Supplemental Material (SM) section. M1 is the model used in \cite{Jedamzik:2013gua,Jedamzik:2018itu}, while M2 was designed to show that the impact on recombination can be weaker despite the PDF having the same second moment $b$. In M2, only a tiny fraction of the total volume is in high density regions, with more of the remaining volume at densities close to the average. Models M1 and M2 approximately bracket the possibilities of a large number of three-zone models with the same average baryon density and clumping factor that we tried.

Having modified CAMB, we use CosmoMC \cite{Lewis:2002ah} to generate Markov chains and find the marginalized posterior distributions for the cosmological parameters in the presence of baryon clumping. We use the Planck 2018 temperature, polarization and CMB lensing spectra ({\textsc\mbox{TT,TE,EE+lowE+lensing}} of \cite{Aghanim:2018eyx}) hereafter called ``Planck'', and the three determinations of $H_0$ by SH0ES, MCP and H0LiCOW referred to as ``H3''.

Fig.~\ref{fig:clumping} shows the marginalized posterior distributions for $b$ and $H_0$. Whereas we find no preferred clumping when using only the CMB data, as in Ref.~\cite{Jedamzik:2018itu} (see the SM section), one can see that after including the three $H_0$ determinations into the analysis the marginalized posterior probability clearly prefers clumping of the order $\sim 0.5$. Moreover, as expected, due to the decreased sound horizon, the preferred value of the Hubble constant is larger and in better agreement with the $H_0$ observations. A further increase in clumping 
seems to be ruled out by the CMB data as it probably 
results in unacceptably large distortions of the 
Silk damping tail, precluding higher values of the inferred $H_0$. While zero clumping is essentially ruled out for both M1 and M2, Fig.~\ref{fig:clumping} demonstrates the fairly large dependence of the effect on the yet unknown baryon density PDF. A more detailed investigation is under way.

\begin{figure*}[!tbph]
\centering
\includegraphics[width=0.48\textwidth]{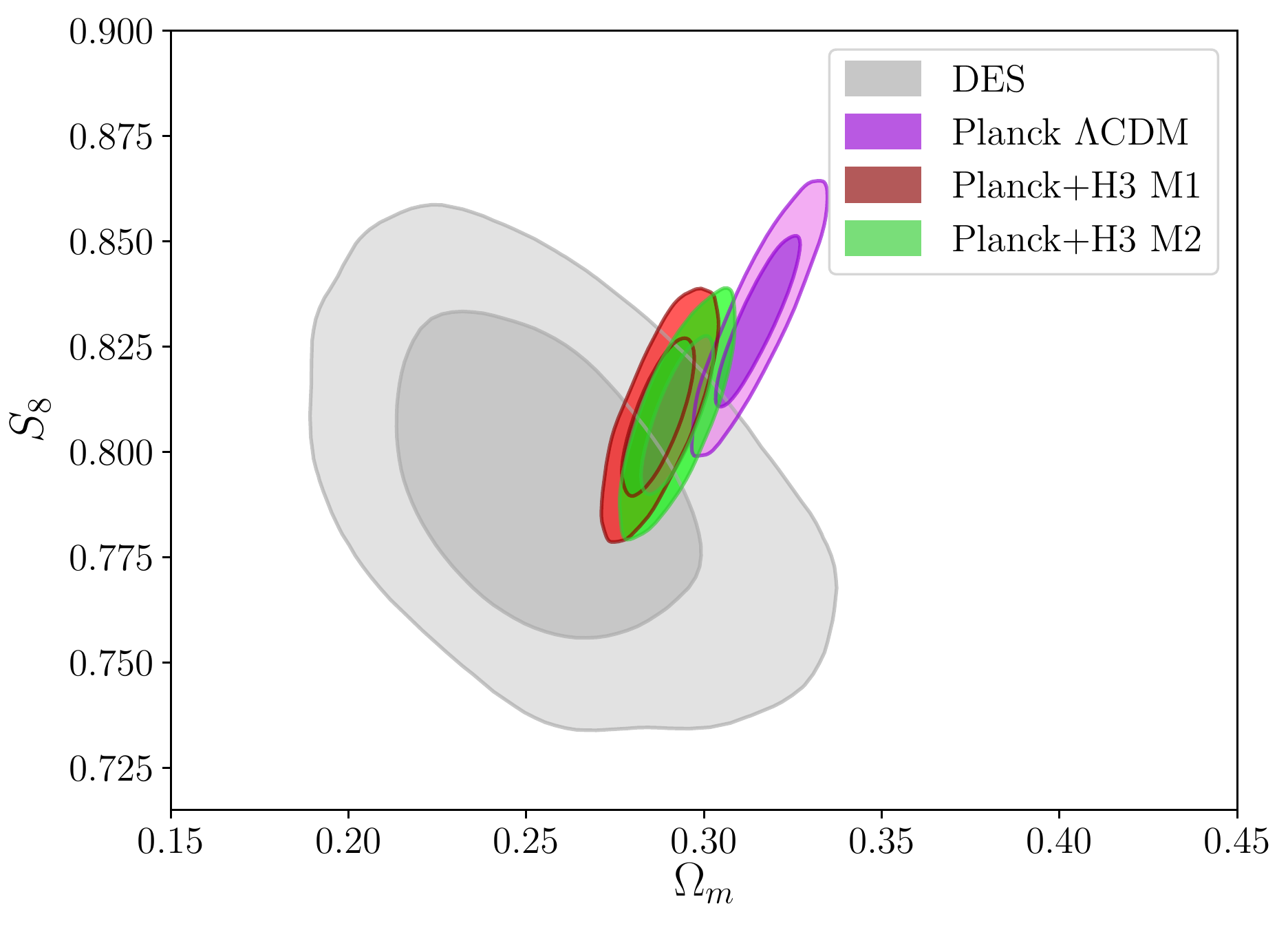} \includegraphics[width=0.48\textwidth]{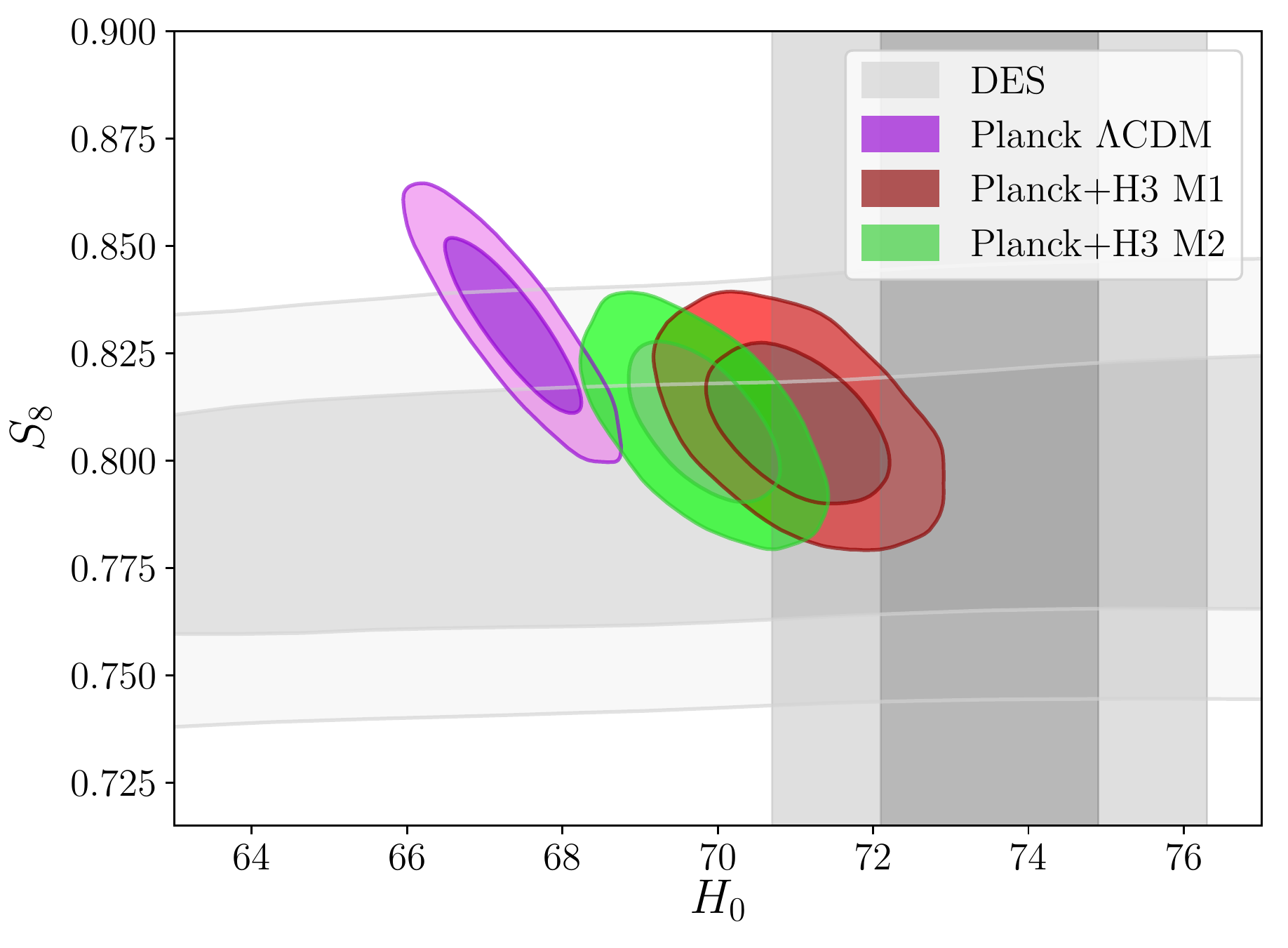}
\caption{\label{fig:S8} The marginalized PDF contours for $S_8-\Omega_m$ (left) showing the mild tension between Planck and DES Y1 when interpreted within $\Lambda$CDM and its resolution after accounting for clumping. The right panel shows how M1 and M2 simultaneously alleviate the $H_0$ and $S_8$ tensions. The shaded vertical regions show the $68$ and $95$\% CL bands of $H_0$ from SH0ES \cite{Reid:2019tiq} }
\end{figure*} 

Further results of the effects of clumping on current cosmological tensions may be observed in Fig.~\ref{fig:S8}. The M1 (M2) clumping model prefer $\Omega_m = 0.2873 \pm 0.0064$ ($0.2926\pm 0.0064$) and $S_8=0.809 \pm 0.012$ ($0.809 \pm 0.012$), significantly lower than in the Planck best fit $\Lambda$CDM model and in good agreement with the values determined from the DES-Y1 weak lensing and galaxy clustering data independent of recombination physics. The possible resolution of both tensions by one well-motivated physical addition to $\Lambda$CDM, small-scale baryon clumping, is graphically presented in the right panel of Fig.~\ref{fig:S8}. Depending on the baryon PDF, the Hubble constant tension is reduced from $4.2\sigma$ to $\lesssim 2\sigma$ ($\lesssim 3\sigma$) for M1 (M2), whereas the $S_8-\Omega_m$ tension is removed. 

How does the addition of baryon clumping affect the goodness of fit to the Planck CMB data? One can see in Fig.~\ref{sup_fig:clumping} of the SM section that without allowing for clumping $\Lambda$CDM prefers a lower value of $H_0$ even after adding the $H3$ data. The statistical weight of the CMB dataset, which is both very large and very precise, is much higher than that of the three $H3$ points and adding the latter has limited impact on cosmological parameters. Allowing for clumping using Model 1 makes the decisive difference, moving the best fit to $H_0=71.03 \pm 0.74$ km s$^{-1}$ Mpc$^{-1}$ and a non-zero clumping value of $b=0.61^{+0.16}_{-0.20}$ at 68\% CL ($^{+0.35}_{-0.33}$ at 95\% CL), while Model 2 gives $H_0=69.81 \pm 0.62$ km s$^{-1}$ Mpc$^{-1}$ and $b=0.31 \pm 0.11$ at 68\% CL ($\pm 0.22$ at 95\% CL) -- a $4\sigma$ ($3\sigma$) detection of clumping in M1 (M2). The mean $\chi^2$ of the $l>29$ TT, TE, and EE binned multipoles portion (the ``Plik'' part) for the Planck+H3 best-fit M1 model is larger than that of the Planck $\Lambda$CDM by $6.7$, which is comparable to the $1\sigma$ uncertainty in $\chi^2_{\rm plik}$ (see Table~\ref{tab:params} in the SM section). This means that Planck+H3 M1 is essentially as good a fit to CMB as the Planck $\Lambda$CDM. A good statistical measure for judging the goodness-of-fit of a particular model to the CMB data is the probability-to-exceed (PTE). Assuming the model is correct, the PTE quantifies the probability of statistical fluctuations in the data resulting in a worse fit. Taking only the Plik likelihood, the Planck collaboration finds PTE $\approx 0.2$ for their best-fit model (Table~20 of \cite{Aghanim:2019ame}). This drops to a PTE of $\approx 0.17 - 0.16$ for both the Planck+H3 best-fit M1 and M2, hardly a serious degradation. However, it is the latter models that alleviate the two existing tensions. The change in $\chi^2$ for the Planck lower non-Gaussian multipoles and CMB lensing is minor.

The impact of including additional datasets is discussed in the SM section. In particular, adding the BAO data\footnote{We treat the BAO data with caution until the potential bias due to using a $\Lambda$CDM based template in BAO likelihoods is estimated for our class of models. During the pre-submission stages of our paper a relevant study examining this issue for several ``exotic'' alternatives to $\Lambda$CDM appeared online \cite{Bernal:2004}, finding that the bias is negligible for the models they considered. Confirming this for baryon clumping is left for a future study.} tends to reduce the value of $H_0$ to $\sim 70.5$ km s$^{-1}$ Mpc$^{-1}$ for M1. This is because the same shift in the sound horizon requires a smaller adjustment of $H_0$ to preserve the angular scale of the acoustic feature measured at low redshifts of $z \sim 0.5$ compared to that at $z \sim 1000$. We note that adding the new parameter $b$ results in only minor changes to the tightness of the posterior distributions of $H_0$ and other $\Lambda$CDM parameters (see the SM section). This shows that clumping really solves the tensions, as opposed to simply allowing more parameter freedom to accommodate independent datasets.

We have so far assumed that the mildly non-linear clumping is due to the existence of PMFs. Are there alternative sources of baryon clumping? An excess of small-scale power of inflationary adiabatic perturbations would be erased by Silk damping. The situation is different for small-scale isocurvature baryon fluctuations that survive Silk damping but are, nevertheless, constrained by Big Bang nucleosynthesis \cite{Jedamzik:1994ux} (BBN). Taking the precise inferences of the primordial deuterium abundance from quasar absorption line systems at face value \cite{Cooke:2016rky,Balashev:2015hoe}, even mildly non-linear isocurvature fluctuations should be ruled out. PMFs have the advantage that they survive Silk damping \cite{Jedamzik:1996wp,Subramanian:1997gi} and actively source the generation of inhomogeneities only shortly before recombination and not during the BBN. The same may apply for cosmic string loops or accreting PBHs, however, it is questionable if they can produce the large volume filling baryon clumping as observed with PMFs.

Magnetic fields are ubiquitous in the universe, observed essentially in all astrophysical environments, including $\sim$ micro-Gauss ($\mu$G) fields in galaxies and clusters of galaxies. Whereas magnetic fields coherent on galactic scales are often believed to result from the dynamo amplification of a pre-existing seed field \cite{Subramanian:2019jyd}, such as that produced in shocks during the collapse of the galaxy, the origin of cluster magnetic fields could possibly be explained by an interplay of galactic dynamos and outflows. More difficult is the explanation of $\sim \mu$G fields in proto-galaxies too young to have gone through the number of revolutions necessary for the dynamo to work \cite{Athreya:1998}. There is also indirect evidence from observations of $\gamma$-rays from TeV-blazars for the existence of an essentially volume filling magnetic field in cosmic voids \cite{Neronov:1900zz,Tashiro:2013ita,Chen:2014rsa}. Since the field strength is likely weak, such fields may possibly be explained by the combined action of outflows from many galaxies. However, none of these astrophysical explanations are well-understood, or have been explicitly shown to work. A magnetic field capable of generating baryon density fluctuations with $b \approx 0.5$ corresponds to a pre-recombination PMF of $\sim 0.07\,$nG comoving strength~\cite{Jedamzik:2018itu}. The PMF that survives to low redshifts may still be a factor $6$ less (for non-helical fields), depending on the magnetogenesis scenario, resulting in $\sim 0.01\,$nG pre-structure formation fields (see discussion below). Such field strength are just of the right order to explain the cluster fields, which require a pre-collapse magnetic field of $\sim 0.005\,$nG irrespective of its coherence length \cite{Banerjee:2003xk,Dolag:99,Dolag:02}. Thus, a discovery of PMFs at recombination of this strength would have the stunning byproduct of explaining entirely the galactic and cluster magnetic fields, and the fields in the extragalactic medium.

PMFs could have been generated during cosmological first order phase transitions in the early universe \cite{Vachaspati:1991nm}, during inflation \cite{Turner:1987bw,Ratra:1991bn}, or at the end of inflation \cite{DiazGil:2007dy} (see \cite{Durrer:2013pga} for a review). A detection of the PMF would offer an invaluable insight into the physics of the early universe \cite{Subramanian:2015lua,Vachaspati:2016xji}. Inflationary models of magnetogenesis \cite{Turner:1987bw,Ratra:1991bn}, by nature, have to result in an approximately scale-invariant PMF spectrum not to be ruled out. PMFs generated via causal processes during phase transitions always develop a unique blue Batchelor spectra with most power on very small scales \cite{Durrer:2003ja,Saveliev:2012ea}. Once produced, their subsequent evolution is much more dramatic than that of the inflationary magnetic fields \cite{Banerjee:2004df,Campanelli:2007tc,Campanelli:2013iaa}. Due to the high conductivity in the early universe, magnetic helicity is essentially conserved, whereas magnetic energy gets dissipated. A causally produced PMF of considerable strength will dissipate many orders of magnitude of its total initial energy density prior to recombination, and a factor of $\sim 20-40$ during/after recombination, depending on the helicity. In general, fields of even a small initial magnetic helicity will develop to be maximally helical during the course of the evolution and decay slower afterwards compared to their non-helical counterparts. It is generally believed that the cosmological electroweak transition, in best case, may only produce pre-recombination fields of $\sim 0.1\,$nG when some initial helicity is present \cite{Wagstaff:2014fla}. Interestingly, non-zero helicity has been linked to the possible generation of the baryon asymmetry during the electroweak transition \cite{Vachaspati:2001nb}, although the predicted helicity, in units of the maximal one, can be at most $h_m\sim 10^{-24}$ \cite{Wagstaff:2014fla}. However, a much larger helicity of $h_m\sim 10^{-3}$ is required to achieve $\sim 0.1\,$nG fields before recombination (see Fig.~19 of \cite{Banerjee:2004df}). In this context, new developments in chiral MHD linking helicity to left-right-handed particle asymmetries are worth noting~\cite{Boyarsky:2011uy,Schober:2018wlo}. On the other hand, the mere presence of PMFs during the electroweak transition has impact on the efficiency of baryogenesis itself~\cite{Comelli:1999gt,DeSimone:2011ek}. Last, but not least, the conclusions regarding the requirements to produce $\sim 0.1\,$nG fields may be changed by the possible discovery of magnetic inverse cascades in non-helical magnetic fields \cite{Brandenburg:2014mwa,Zrake:2014mta}. 

It is our belief that amending $\Lambda$CDM by clumping due to magnetic fields is a very modest and physically reasonable extension that shows promise to resolve the existing tensions. Ultimately, the existence of PMFs may have to be established by further smoking guns in future observations. For causally produced fields, an initial estimate shows that a future mission like the Primordial Inflation Explorer (PIXIE)~\cite{Kogut_2011}, targeting the spectral distortions of CMB, is sensitive enough to detect the dissipation of magnetic fields at redshifts $z\sim 10^4$. A competing effect may be distortions in the spectrum due to Silk damping~\cite{Chluba:2019nxa}. The metric fluctuations induced by causal PMFs do not make a detectable contribution to anisotropies in the CMB, while a scale-invariant PMF can only make a detectable impact if the field is of $\sim 1$ nG strength \cite{Seshadri:2000ky,Mack:2001gc,Lewis:2004ef,Finelli:2008xh,Shaw:2009nf,Ade:2015cva,Zucca:2016iur} -- at least an order of magnitude stronger than that required to produce a detectable clumping signature. The next generation CMB polarization experiments, such as Probe of Inflation and Cosmic Origins (PICO) \cite{Hanany:2019lle} and CMB-S4 \cite{Abazajian:2019eic}, can probe scale-invariant PMFs of $\sim 0.1$ nG strength through measurements of the Faraday Rotation induced at the epoch of last scattering \cite{Pogosian:2019jbt}. Additional constraints on large scale magnetic fields come from observations of cosmic rays \cite{Archambault:2017hvo,Tiede:2017aql}.

In this {\it Letter}, we have shown that mildly non-linear, small-scale baryon inhomogeneities in the universe existing before recombination, as expected to naturally emerge from $\sim 0.1$ nG pre-recombination PMFs \cite{Jedamzik:2013gua,Jedamzik:2018itu}, show promise to resolve the Hubble tension. Such scenarios result in an inevitable reduction of the sound horizon at recombination, and thereby may bring local measurements of $H_0$ in agreement with the inferences from CMB. Interestingly, if PMFs of such strength existed, the origin of galactic, cluster, and extragalactic magnetic fields would be explained. 

{\it Acknowledgments.} We thank Antony Lewis, Simone Peirone, Vivian Poulin, Marco Raveri, Andrey Saveliev, Meir Shimon, Tanmay Vachaspati, Alex Zucca and Gong-Bo Zhao for valuable assistance and discussions. We gratefully acknowledge using GetDist \cite{Lewis:2019xzd}. This research was enabled in part by support provided by SciNet (www.scinethpc.ca) and Compute Canada (www.computecanada.ca). L.P. is supported in part by the National Sciences and Engineering Research Council (NSERC) of Canada.

\appendix
\section{Supplemental Material}


{\it The three-zone model.} Deriving the baryon density PDFs requires many MHD simulations for different magnetic field strengths, spectral indices, and helicities. Furthermore, the entire evolution of the PDF before recombination has to be known, although we expect the main trend to emerge even when assuming a non-evolving PDF. We leave this numerically expensive study for a future publication and, in the meantime, use a simple three-zone model.  The model is described by the density parameters $\Delta_i$ and volume fractions $f_V^i$ in each zone. Baryon densities in the individual zones are simply given by $n_b^i = \langle n_b\rangle \Delta_i$. Parameters $\Delta_i$ and $f_V^i$ have to fulfil the following constraints:
\begin{eqnarray}
\sum_{i=1}^3 f_V^i =1, \ \sum_{i=1}^3 f_V^i\Delta_i = 1, \ 
\sum_{i=1}^3 f_V^i\Delta_i^2 = 1 + b\, ,
\end{eqnarray}
{\it i.e.} the total volume fraction is one and the three-zone model has average density $\langle n_b\rangle$ and clumping factor $b$.
This leads to three constraints for six free parameters ${f_V^i,\Delta_i}$, such that one may choose three parameters freely. We have chosen $f_V^2 = 1/3, \Delta_1 = 0.1, \Delta_2 = 1$ for M1 and $f_V^2 = 1/3, \Delta_1 = 0.3, \Delta_2 = 1$ for M2. To obtain the average ionization fraction $\langle \chi_e \rangle$, we compute the ionization fraction in each of the zones and take the average, {\it i.e.}
\begin{equation}
\langle \chi_e \rangle = \sum_{i=1}^3 f_V^i\Delta_i \chi_e^i \ .
\end{equation}
In our Monte-Carlo simulations we use a flat prior for the clumping factor
between zero and six.  


\begin{figure}[!tbp]
\centering
\includegraphics[width=0.48\textwidth]{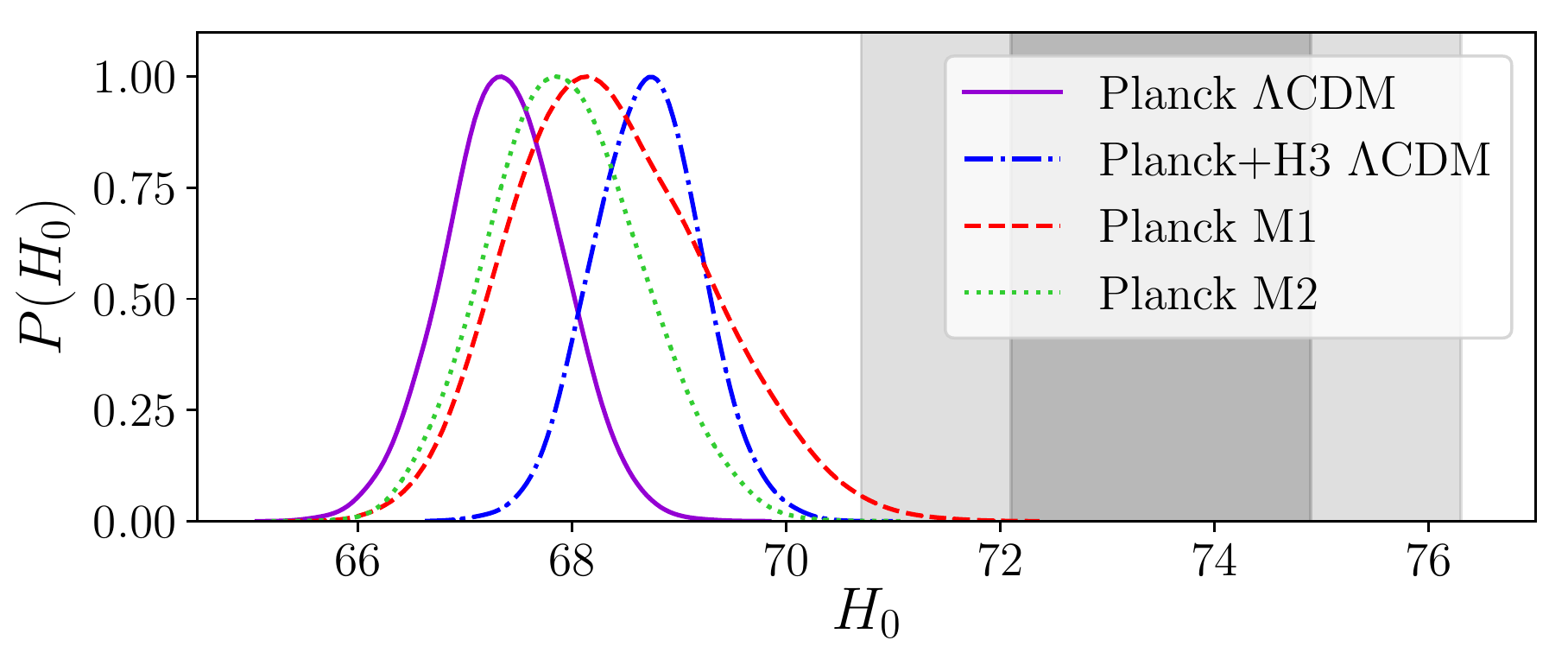}
\includegraphics[width=0.48\textwidth]{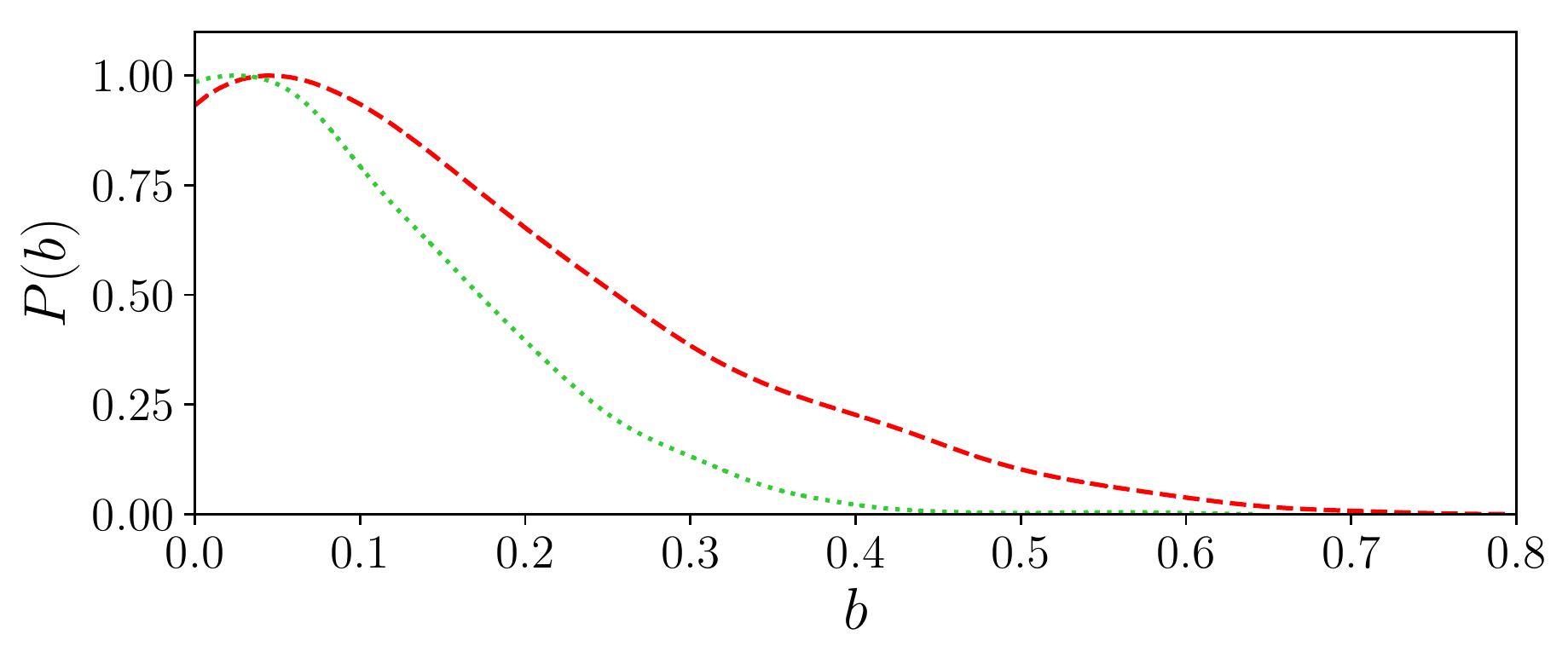}
\caption{\label{sup_fig:clumping} The marginalized posterior PDF of $H_0$ for the Planck best fit $\Lambda$CDM model, the $\Lambda$CDM fit to Planck+H3 and the two clumping models, M1 and M2, fit to just Planck (without H3). The bottom panel shows the PDF of the clumping parameter $b$ in the latter two cases. The shaded regions show the $68$ and $95$\% CL regions of the SH0ES measurement of $H_0$.}
\end{figure}

\begin{table*}[!tbp]
\centering
\begin{tabular}{c|c|c|c|c|c|c}
 & Planck $\Lambda$CDM & Planck+H3 $\Lambda$CDM & Planck+H3 M1 &  Planck+H3 M2  \\
\hline \hline
$\Omega_b h^2$ & $0.02237\pm 0.00015$ & $0.02263\pm 0.00014$ & $0.02270^{+0.00014}_{-0.00016}$ & $0.02280\pm 0.00016$ \\
$\Omega_c h^2$ & $0.1200\pm 0.0012$ & $0.1172\pm 0.0011$ & $0.1216\pm 0.0014$ & $0.1191\pm 0.0012$ \\
$\tau$ & $0.0546\pm 0.0075$  &  $0.0629^{+0.0075}_{-0.0087}$ & $0.0555\pm 0.0073$ & $0.0607^{+0.0071}_{-0.0085}$  \\
$n_s$ & $0.9651\pm 0.0041$ &  $0.9721\pm 0.0040$ & $0.9628\pm 0.0040$ & $0.9734\pm 0.0042$  \\
$b^{(a)}$ & - & - &   $0.61^{+0.16(0.35)(0.57)}_{-0.20(0.33)(0.42)}$ & $0.30\pm 0.11(0.22)(0.34)$  \\
$H_0$ & $67.37\pm 0.54$ & $68.70\pm 0.50$ & $71.03\pm 0.74$ & $69.81\pm 0.62 $  \\
$\Omega_m$ & $0.3151\pm 0.0074$ &$0.2977\pm 0.0064$ & $0.2873\pm 0.0064$ & $0.2926\pm 0.0064$  \\
$\sigma_8$ & $0.8113\pm 0.0060$ & $0.8080\pm 0.0064$ & $0.8265\pm 0.0079$ & $0.8192\pm 0.0075$ \\
$S_8$ & $0.831\pm 0.013$ & $0.805\pm 0.012$ & $0.809\pm 0.012$ & $0.809\pm 0.012 $  \\
$z_*$ & $1089.91\pm 0.26$ & $1089.35\pm 0.24$ &  $1107.9^{+4.2}_{-3.6}$ & $1096.8^{+2.6}_{-2.0} $   \\
$r_*$ & $144.44\pm 0.27$ & $144.96\pm 0.25$ & $142.22\pm 0.65$ & $143.69\pm 0.48 $   \\
$z_{\rm{drag}}$ & $1059.94\pm 0.30$ &  $1060.33\pm 0.29$ & $1076.9^{+3.8}_{-3.4}$ & $1067.4^{+2.4}_{-2.0} $  \\ 
$r_{\rm{drag}}$ & $147.10\pm 0.27$ &  $147.55\pm 0.25$ & $144.89\pm 0.64$ & $146.28\pm 0.49  $ \\
$r_{\rm drag} h$ & $99.11 \pm 0.93$ & $101.36\pm 0.87$ & $102.91\pm 0.92$ & $102.11\pm 0.89 $  \\
$\chi^2_{\rm lensing}$ & $9.23\pm 0.70$ (8.73) & $9.6\pm 1.2$ (8.74) & $9.20\pm 0.66$ (8.91) & $9.33\pm 0.80$ (9.39)  \\
$\chi^2_{\rm plik}$ & $2359.5\pm 6.2$ (2347.6) & $2364.0\pm 6.6$ (2350.93)  & $2366.2\pm 6.7$ (2355.6) & $2367.4\pm 7.1$ (2359.2)  \\
$\chi^2_{\rm lowl}$ & $23.40\pm 0.86$  (23.18) &  $22.36\pm 0.72$ (22.76) & $24.30\pm 0.97$ (24.0) & $22.37\pm 0.72$ (21.9)  \\
$\chi^2_{\rm simall}$ & $397.0\pm 1.8$ (396.0) &   $399.0\pm 3.3$ (397.2) & $397.0\pm 1.7$ (395.6) & $398.2\pm 2.7$ (396.3)  \\
$\chi^2_{\rm prior}$ & $11.6\pm 4.6$ (4.46) & $11.6\pm 4.6$ (4.38) & $11.6\pm 4.5$ (4.21) & $11.9\pm 4.6$ (3.42)  \\
$\chi^2_{\rm CMB}$ & $2789.1 \pm 6.4$ (2775.5) & $2794.9\pm 7.2$ (2779.7) & $2796.8\pm 6.9$ (2784.2) & $2797.3\pm 7.3$ (2786.8)  \\
$\chi^2_{\rm H3}$ & - &  $22\pm 4$ (24.92) & $6.1\pm 3.4$ (5.74) & $12.9\pm 4.2$ (9.62)  \\
$\chi^{2 \rm{(tot)}}_{\rm best fit}$ & $2779.9$ & $2809.0$ & $2794.1$ & $2799.9$ \\
 \hline
\end{tabular}
\caption{\label{tab:params} The mean values and $68$\% CL intervals for the relevant cosmological parameters and $\chi^2$ for $\Lambda$CDM, M1 and M2 fit to a combination of Planck and H3, along with the Planck best fit $\Lambda$CDM. The best fit $\chi^2$ values are shown in parenthesis. {\it $^{(a)}$We show the 95\% and 99.7\% CL uncertainties for the clumping parameter $b$ in parenthesis.}
}
\end{table*}

\begin{table*}[!tbp]
\centering
\begin{tabular}{c|c|c|c|c|c|c}
 & Global $\Lambda$CDM & Global M1 &  Global M2  \\
\hline \hline
$\Omega_b h^2$ & $0.02266\pm 0.00013$ & $0.02266\pm 0.00014        $& $0.02278\pm 0.00015        $\\
$\Omega_c h^2$ & $0.11680\pm 0.00079$ & $0.1210\pm 0.0015          $& $0.1187\pm 0.0010          $\\
$\tau$ & $0.0622^{+0.0070}_{-0.0083}$ & $0.0541\pm 0.0074          $& $0.0587^{+0.0069}_{-0.0077}$\\
$n_s$ & $0.9725\pm 0.0036$ & $0.9624\pm 0.0038          $& $0.9725\pm 0.0037          $\\
$b^{(a)}$ & - & $0.48^{+0.16}_{-0.18}      $& $0.246^{+0.097}_{-0.11}    $\\
$H_0$ & $68.86\pm 0.37$ & $70.57\pm 0.61             $& $69.69\pm 0.47             $\\
$\Omega_m$ & $0.2955\pm 0.0046$ & $0.2898\pm 0.0046          $ & $0.2927\pm 0.0045          $\\
$\sigma_8$ & $0.8056\pm 0.0061$ & $0.8208\pm 0.0078          $& $0.8148\pm 0.0070          $\\
$S_8$ & $0.7995\pm 0.0088$ & $0.8067\pm 0.0094          $& $0.8048\pm 0.0090          $\\
$z_*$ & $1089.28\pm 0.19$ &$1105.0^{+4.6}_{-3.8}      $& $1095.8^{+2.5}_{-2.1}      $\\
$r_*$ & $145.04\pm 0.20$ & $142.65^{+0.65}_{-0.76}    $& $143.89\pm 0.46            $\\
$z_{\rm{drag}}$ & $1060.37\pm 0.29$& $1074.3^{+4.1}_{-3.5}      $& $1066.5^{+2.3}_{-2.0}      $\\
$r_{\rm{drag}}$ & $147.62\pm 0.21$ & $145.31^{+0.65}_{-0.75}    $& $146.48\pm 0.47            $\\
$r_{\rm drag} h$ & $101.66\pm 0.62$ & $102.54\pm 0.66            $& $102.08\pm 0.63            $\\
$\chi^2_{\rm lensing}      $ & $9.9\pm 1.3 (9.1)$ & $9.17\pm 0.67 (8.6)$& $9.48\pm 0.94 (8.7)$\\
$\chi^2_{\rm plik}         $ & $2365.4\pm 6.4 (2354.8)$ & $2364.6\pm 6.2 (2353.5)$& $2366.8\pm 6.5 (2352.9)$\\
$\chi^2_{\rm lowl}         $ & $22.27\pm 0.66 (22.6) $& $24.26\pm 0.96 (25.1) $& $22.43\pm 0.69 (22.5)$\\
$\chi^2_{\rm simall}       $ & $398.6\pm 2.9 (398.3)$& $396.8\pm 1.6 (395.7)$& $397.6\pm 2.2 (399.2)$\\
$\chi^2_{\rm H3}      $ & $20.0\pm 3.1 (23.1)$& $8.3\pm 3.3 (6.92)$& $13.6\pm 3.3 (13.5)$\\
$\chi^2_{\rm SN}          $ & $1034.80\pm 0.09 (1034.73)$& $1034.96\pm 0.19 (1034.92) $& $1034.86\pm 0.14 (1034.8)$\\
$\chi^2_{\rm 6DF}          $ & $0.073\pm 0.067 (0.013)$& $0.20\pm 0.12 (0.18) $& $0.123\pm 0.089 (0.09) $\\
$\chi^2_{\rm MGS}          $ & $2.51\pm 0.43 (2.11)$& $3.14\pm 0.48 (3.18)$& $2.81\pm 0.45 (2.76)$\\
$\chi^2_{\rm DR14LYA}      $ & $4.00\pm 0.21 (4.15) $& $3.81\pm 0.21 (3.79)  $& $3.90\pm 0.21 (3.93)$\\
$\chi^2_{\rm DR12BAO}      $ & $3.99\pm 0.71 (3.42)$& $5.4\pm 1.5 (5.16) $& $4.5\pm 1.0 (4.11)$\\
$\chi^2_{\rm DES}          $ & $516.4\pm 4.4 (512.1)$& $516.8\pm 4.6 (512.9) $& $517.1\pm 4.6 (511.1)$\\
$\chi^2_{\rm prior}        $ & $24\pm 7 (10.1)$& $24\pm 7 (13.4)$ & $24\pm 7 (12.0)$\\
$\chi^2_{\rm CMB}          $ & $2796.2\pm 6.9 (2784.9) $& $2794.9\pm 6.4 (2782.89)$ & $2796.3\pm 6.6 (2783.2)$\\
$\chi^2_{\rm BAO}          $ & $10.58\pm 0.98 (9.7) $& $12.6\pm 1.9 (12.3)$& $11.3\pm 1.4 (10.9)$\\
$\chi^{2 \rm{(tot)}}_{\rm best fit}$ & $4374.6$ & $4363.3$ & $4365.6$\\
 \hline
\end{tabular}
\caption{\label{tab:params_global} Same as in Table~\ref{tab:params} but for the global fit to a combination of Planck, BAO, SN, DES and H3.}
\end{table*}

{\it Differences with the analysis of Ref.~\cite{Jedamzik:2018itu}.} 
Ref.~\cite{Jedamzik:2018itu} finds an upper limit on post-recombination 
fields of $8.9$pG and $47$pG at $95\%$ confidence level for causally produced
and scale-invariant fields, respectively. These limits are derived from
a combination of the complete 2013 Planck data, WMAP polarization data,
Hubble constant determination from HST, $H_0 = 73.8 \pm 2.4$ km/s/Mpc,
a variety of BAO data, supernovae
data, as well as structure formation data from SDSS. Note that erroneously
in Ref.~\cite{Jedamzik:2018itu} it was written that the limit is only due
to Planck and WMAP data.  
At first glance these limits seem to be in stark conflict with the findings here. However, importantly, it should be noted that the field strengths quoted in Ref.~\cite{Jedamzik:2018itu} are post-recombination, whereas the strength
noted in the current paper are pre-recombination. A causally produced non-helical field of $\sim 9$pG corresponds to a pre-recombination field of $\sim 50$pG, close to the pre-recombination fields needed to explain the
possible detection of clumping here. Limits reported in Ref.~\cite{Jedamzik:2018itu} certainly have to be revised, given the results 
here. However, these limits are expected to change at most by a factor of two, 
as the clumping factor scales as a high power of magnetic field strength.

{\it Additional fits and cosmological parameters.}
Here we present additional results that aid the interpretation of our main results. Fig.~\ref{sup_fig:clumping} shows three additional posterior PDFs of $H_0$ to compare to those in Fig.~\ref{fig:clumping}, along with corresponding PDFs of the clumping factor $b$, while Tables~\ref{tab:params} and \ref{tab:params_global} list the bounds on the relevant parameters and the $\chi^2$ of the data likelihoods used in the analysis. 

One can see from Fig.~\ref{sup_fig:clumping} (as well as the Table) that simply adding the H3 data to Planck results in a limited shift of the best fit $H_0$ --- the mean value changes from 67.4 to 68.7. This comes at the cost of a significant increase in the total $\chi^2$ ($+29$ after adding 3 data points) compared to fitting $\Lambda$CDM to Planck alone. Fig.~\ref{sup_fig:clumping} also shows that fitting the clumping models M1 and M2 to Planck data alone does not result in a detection of clumping, although their posterior distributions are sufficiently broad to be consistent with a non-zero $b$. We also see that allowing for clumping results in slightly broadened PDFs for $H_0$ along with preference for higher values. This reflects the fact that there are models with higher values of $H_0$ and non-zero $b$ that give acceptable fits to the Planck data. Adding the H3 data narrows the choice of the best fit clumping model down, slimming the PDFs for $H_0$ and providing a clear detection of $b$, as seen in Fig.~\ref{fig:clumping}.

Several additional insights can be made by examining Table~\ref{tab:params}. One can see that, while the Planck best fit $\Lambda$CDM has the lowest $\chi^2_{plik}$ (corresponding to the high-$\ell$ TT, TE, EE Planck spectra), the Planck+H3 best fit M1 model has a mean $\chi^2_{plik}$ that is larger by just one standard deviation. The differences in $\chi^2$ for the other three Planck likelihoods are essentially none. Given that the Planck data has known $\sim 1\sigma$-level internal inconsistencies when interpreted in the context of the $\Lambda$CDM model \cite{Aghanim:2018eyx}, one should not assign significance to $\sim 1\sigma$ differences in $\chi^2$. One can use criteria such as the PTE to assess the goodness of the fit to the Planck data and whether a model simultaneously fits multiple datasets well. One can also note that the parameter uncertainties remain small after adding the additional parameter $b$.

Table~\ref{tab:params_global} compares the ``global'' fits of LCDM, M1 and M2 to Planck and H3 combined with the BAO data from the 6dF Galaxy Survey \cite{Beutler_2011}, the SDSS DR7 Main Galaxy Sample (MGS) \cite{Ross:2014qpa}, the BOSS DR12 Consensus BAO \cite{Alam:2016hwk} and the cross-correlation of Ly$\alpha$ absorption and quasars in eBOSS DR14 \cite{Blomqvist:2019rah}, along with the Pantheon supernovae sample \cite{Scolnic:2017caz} and the DES-Y1 weak lensing and galaxy clustering data \cite{Abbott:2017wau}. We note that the mean values of $H_0$ in M1 and M2 are smaller than those obtained by fitting to Planck+H3. This is due to the inclusion of the BAO data which requires a smaller increase in $H_0$ compared to CMB to accommodate the same shift in the value of the sound horizon. A detailed study of the influence of magnetic fields on the amplitude and the shape of the BAO peaks using extensive MHD simulations is currently in progress.

\end{document}